# Short-range ordering of heavy element columns in nickel based superalloys

Yushi Luo[1], Lihui Zhang[1], Yumei Wang[2], Binghui Ge[2], Wei Guo[3], Jie Zhan[4], Jianxin zhang[5], Jing Zhu[6],

1 National Key Laboratory of Science and Technology on Advanced High Temperature Structural Materials, Beijing Institute of Aeronautical Materials, Beijing 100095, People's Republic of China
2 Beijing National Laboratory for Condensed Matter Physics,
Institute of Physics, Chinese Academy of Sciences, Beijing 100190, People's Republic of China;
3 School of Physics, Beijing Institute of Technology, Beijing 100081, People's Republic of China;
4 National Key Laboratory of Crystal Materials, Institute of Crystal Materials,
Shandong University, Jinan 250100, China
5 Department of Materials Science and Engineering, Shandong University, Jinan, 250100, People's Republic of China;
6 Beijing National Center for Electron Microscopy, Laboratory of Advanced Materials, Department of Materials Science and Engineering, Tsinghua University, Beijing 100084, People's Republic of China;

To obtain comprehensive performance, heavy elements were added into superalloys for solid solution hardening. In this article, it is found by scanning transmission electron microscope observation that rather than distribute randomly heavy-atom columns prefer to align along <100> and <110> direction and form a short-range ordering with the heavy-element stripes 1-2 nm in length. Due to the strong bonding strength between the refractory elements and Ni atoms, this short-range ordering will be beneficial to the mechanical performances.

Keyword: Superalloy; Short-range ordering; HAADF; Heavy element

Single crystal nickel-based superalloys have been used as the material for gas turbine blades in the aerospace industry and land-based applications, in which ordered γ′ precipitates with structure L1$_2$ are embedded in a γ matrix with a face-centered cubic (FCC) structure [1]. To improve the mechanical performance, refractory elements such as Ta, W, Mo and Re are added for solid solution hardening. To understand the strengthen mechanism of these refractory elements, their distribution should be determined at first.

In general, Mo and Re were mainly partitioned into the γ phase and Ta mainly into the γ′ phase, while W showed no preferentially partitioning [2, 3]. For the crept superalloys, heavy atoms such as Re[2, 4, 5] and W[5] are found to be enriched in the γ phases close to the γ/γ′ interfaces, and for Re there is enrichment at the tip of the interfacial protrusions [6, 7] and for Re and Mo at the core of dislocations close to the interface [5]. As to Re cluster, no consensus has been reached up to now [8], some arguing for its presence [3, 5, 9-12] and some arguing against its presence [2, 11, 13, 14]. Nevertheless, heavy atom clusters were observed directly by means of electron microscope, high angle annular dark field (HAADF) imaging [5], although their composition has not been determined yet. Recently, a new type of Re ordering was reported in the form of D1$_a$-Ni$_4$Re [15], and even in high temperature a short-range ordering of Re atoms was thought to be left.

In this article, the scanning transmission electron microscopy (STEM), HAADF imaging and energy dispersive X-ray spectroscopy (EDS) techniques have been combined to study the second-generation single crystal superalloy DD6. Heavy-atom columns were found to align along <100> and <110> direction and a short-range of ordering were formed with the heavy-element stripe 1-2 nm in length.

Different from the high-resolution imaging in transmission electron microscopy (TEM), the image contrast of which changes with thickness of samples and imaging focus etc., contrast of HAADF imaging mainly arises from contribution of the thermal diffusion scattering for the large collection angle and follows almost $Z^2$ dependency with respect to the atomic number [16-18], then heavy alloying elements can easily be distinguished, displaying with higher contrast while light elements with low contrast [19].

The composition of DD6 is given in Table 1. Heat treatments were performed according to following regime: 1290 °C/1 h + 1300°C /2 h + 1315 °C / 4 h/AC + 1120 °C /4 h/AC + 870 °C /32 h/AC. Details of DD6 preparation have been described previously [20]. As shown in Table 1, DD6 contains 2 wt.% Re, which is about two-thirds that of other second-generation single crystal superalloys, such as PWA1484 CMSX-4 and René N5. Discs with diameters of 3 mm suitable for STEM observations were punched out of samples and thin foils were prepared by electrochemical polishing and cryo ion milling. STEM observations were carried out in a FEI Titan 80-300 and a FEI F20 electron microscope equipped with an EDS spectrometer.

Table1. Composition of DD6(wt.%).

| Cr  | Co | Mo | W | Ta  | Re | Nb  | Al  | Hf  | Ni      |
|-----|----|----|---|-----|----|-----|-----|-----|---------|
| 4.3 | 9  | 2  | 8 | 7.5 | 2  | 0.5 | 5.6 | 0.1 | Balance |

Figure 1a is the low-magnification HAADF image of the superalloy DD6. For heavy elements mainly partition into the γ phase, it displays with higher contrast than γ′ phases. The area denoted by a white rectangle in Figure 1a was selected to produce element mapping images as shown in Figure 1b. Similar to other reports [1], the alloying elements Al and Ta are found to be mainly partitioned into the γ′ phase, while the elements Cr, Co, Mo and Re mainly partitioned into the γ phase, and W showed no preferentially partitioning.

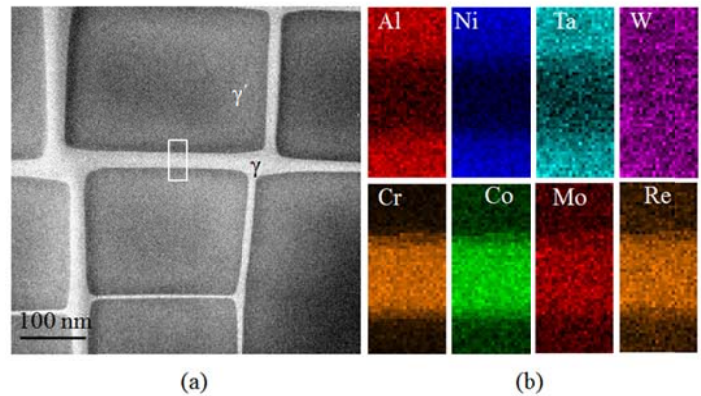

Figure 1. (a) Low-magnification HAADF image of the superalloy DD6 before creep test. (b) Element mapping of major constituents of the area denoted by a white rectangle in (a).

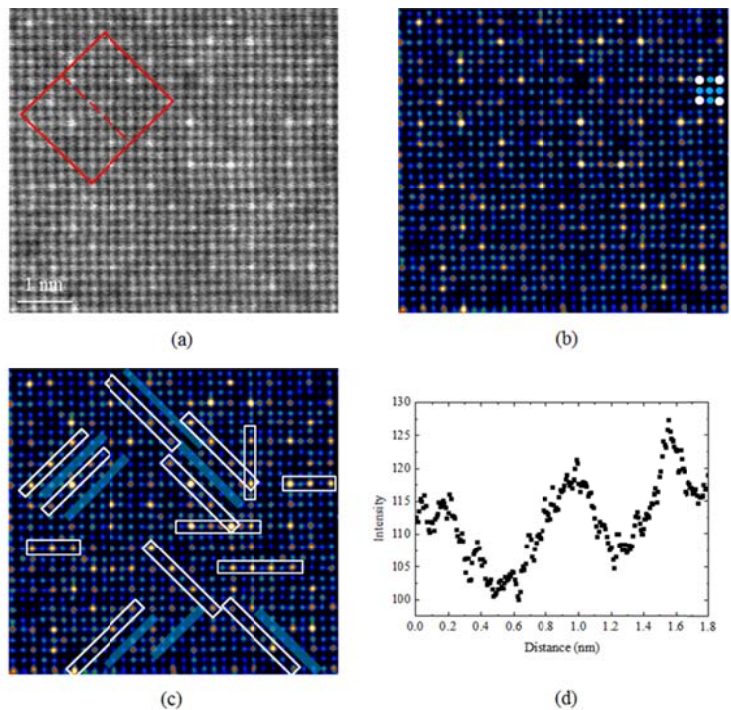

Figure 2. (a) HAADF image of the γ′ phase in [001] direction. (b) and (c) Processed images of (a) after bandpass filtering and displaying in color temperature. The projected structure model of the atoms. White rectangles in (c) indicate the heavy-atom stripes while light blue ones denote the dark areas between the heavy-atom stripes. (d) Intensity profile along the red dashed line in (a), averaged by integration of the area of the red rectangle.

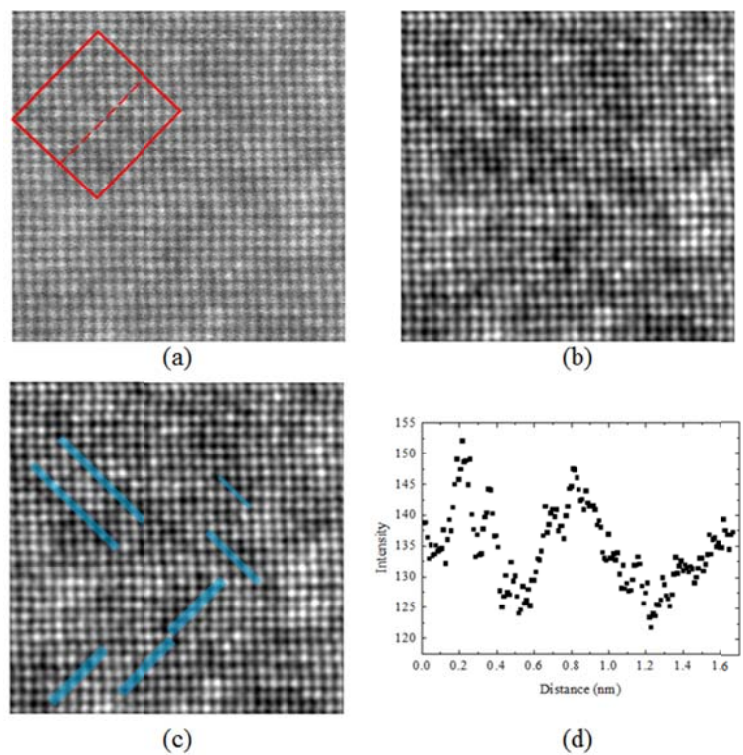

Figure 3. (a) HAADF image of the γ phase in [001] direction. (b) and (c) Processed images of (a) after bandpass filtering. Light blue ones denote the dark areas between the heavy-atom stripes. (d) Intensity profile along the red dashed line in (a), averaged by integration of the area of the red rectangle

Figure 2a is the high-resolution HAADF image of the γ′ phase in [001] zone axis in which bright dots represent heavy atoms, such as Mo, Ta, W and Re. From Figure 2a, heavy atoms are found not to be randomly located, but preferred to be at Al site, which can be seen clearly in Figure 2b, obtained by bandpass filtering on Figure 2a and displaying in color temperature but not in grayscale, and in agreement with the results of the atom probe tomography [21]. The projected model of the γ′ phase is inset on top right with white dots representing Al atoms and blue dots Ni atoms. Moreover, one feature in Figures 2a and b can be found that some heavy atoms align along <100> and <110> direction and white stripes are formed, denoted by a white rectangle in Figure 2c, and the area between the heavy-atom stripes forms the dark stripe, denoted by a light blue rectangle. To illustrate the contrast vibration more clearly, the intensity profile as shown in Figure 2d was made along the red dashed line and averaged by integration of the red rectangle as shown in Figure 2a. Two valleys in

Figure 2d correspond to two dark areas indicated by two blue rectangles in Figure 2c. Similar results of heavy-atom ordering can be observed in the γ phase as shown in Figure 3. Just because more heavy alloying elements partition into the γ phase as shown in Figure 1, the area with less heavy elements can be observed easily as shown by light blue rectangles. Such heavy-atom ordering resembles the results reported by Maisel etc. [15], in which Re is preferred to align in <420> direction in the form of $Ni_4Re$.

In our sample, stripes can be observed not only in (001) plane but also in (100) or (010) plane, so some stripes, if along [011] or [101] direction, will have a non-zero resolved shear stress when under [001] loading. As we know, the final fracture of materials is caused by the breaking of atomic bonds, so the rupture strength can be improved by forming stronger atomic bonds. According to the report that there is stronger bond between refractory elements such as Re, W, Ru, Ta and Mo with Ni atoms [26, 27], then in our case, these kind of heavy-atoms stripe should be of stronger binding. Moreover, as mentioned above the heavy-atom stripes are usually 1-2 nm in length, and distribute in 3 dimensions, then a partial heavy-element stripe network was formed, just like an uncompleted skeleton. Due to these two factors, stronger bond and formation of a partial network, the heavy-element ordering may help to improve the mechanical performances[15], similar with the interfacial dislocation networks [28]. More works need to be done, however, to study this type of stripe, as to which element is preferred, and how to control its length, and so on.

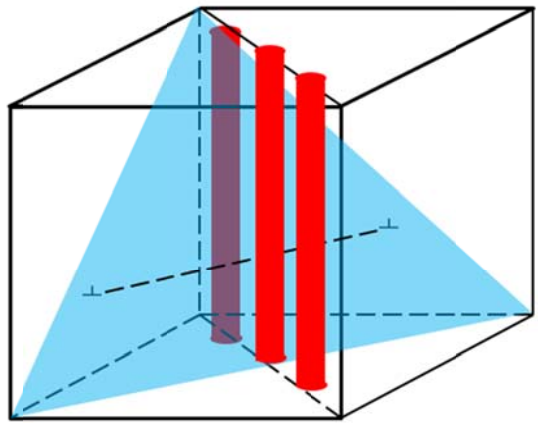

Figure 4. 3D Schematic of hindering motion of dislocation by heavy-atom column stripes. Red cylinders represent heavy atom column. They will hinder the motion of dislocations indicated by '⊥' in {111} slip plane described by light blue rectangle.

Moreover, according to the composition of DD6 as shown Table 1, heavy atoms such as Ta, W and Re are almost 7 atom percent. That is to say, there should be 1 heavy atom within 4 unit cell of $Ni_3Al$. While the thickness of the sample for the Figure 2a is about 20-30 nm determined by the relative log-ratio method on electron energy loss spectra, almost 60 unit cell, such that at every Al position projected in [001] direction there should be about 15 atoms among Ta, W and Re in average. Then bright dots in Figure 2 should correspond to more heavy atoms, and they are actually heavy-atom columns. Thus, the above-mentioned ordering should be a heavy-atom-column ordering, the schematic of which is shown as a group of red cylinders in Figure 4. As we know, for FCC structure, dislocations indicated by ⊢ in Figure 4 move in the {111} plane (blue plane), then the short range of heavy-atom-column ordering as shown in Figure 4 are thought to be more beneficial to hinder the motion of dislocations comparing with just heavy-atom ordering or mere a heavy-atom column.

In conclusion, the second-generation single crystal superalloy DD6 was studied by means of STEM and HAADF imaging. Heavy-atom columns were found to align in the <100> and <110> direction and form a heavy-atom stripe with about 1-2 nm in length. Due to the high diffusion activation energy of heavy atoms and stronger bonding strength of the refractory element with Ni this short-range ordering will help to improve the mechanical performances. The author believes that this work shall stimulate further researches, for example to determine and even tune the composition of heavy atom stripe or form a long range of ordering.

## Acknowledgement

This work was supported by the National Natural Science Foundation of China (Grant number: 11374332, 11474329, 51372138, 51271097). This work made use of the resources of the Beijing National Center for Electron Microscopy at Tsinghua University.